\documentclass[11pt,twoside]{article}
\usepackage{./asp2014}

\aspSuppressVolSlug
\resetcounters

\begin{document}

\title{Supernovae Distribution and Host Galaxy Properties}
\author{A.~A.~Hakobyan,$^1$ A.~G.~Karapetyan,$^1$ L.~V.~Barkhudaryan,$^1$ G.~A.~Mamon,$^2$ D.~Kunth,$^2$
        A.~R.~Petrosian,$^1$ V.~Adibekyan,$^3$ L.~S.~Aramyan,$^1$ and M.~Turatto$^4$
\affil{$^1$Byurakan Astrophysical Observatory, Byurakan, Armenia;
       \email{hakobyan@bao.sci.am}\\
       $^2$Institut d'Astrophysique de Paris, Paris, France;\\
       $^3$Instituto de Astrof\'{i}sica e Ci\^{e}ncia do Espa\c{c}o, Porto, Portugal;\\
       $^4$Osservatorio Astronomico di Padova, Padova, Italy}}

\paperauthor{A.~A.~Hakobyan}{hakobyan@bao.sci.am}{}{Byurakan Astrophysical Observatory}{}{Byurakan}{}{}{Armenia}
\paperauthor{A.~G.~Karapetyan}{}{}{Byurakan Astrophysical Observatory}{}{Byurakan}{}{}{Armenia}
\paperauthor{L.~V.~Barkhudaryan}{}{}{Byurakan Astrophysical Observatory}{}{Byurakan}{}{}{Armenia}
\paperauthor{G.~A.~Mamon}{}{}{Institut d'Astrophysique de Paris}{}{Paris}{}{}{France}
\paperauthor{D.~Kunth}{}{}{Institut d'Astrophysique de Paris}{}{Paris}{}{}{France}
\paperauthor{A.~R.~Petrosian}{}{}{Byurakan Astrophysical Observatory}{}{Byurakan}{}{}{Armenia}
\paperauthor{V.~Adibekyan}{}{}{Instituto de Astrof\'{i}sica e Ci\^{e}ncia do Espa\c{c}o}{}{Porto}{}{}{Portugal}
\paperauthor{L.~S.~Aramyan}{}{}{Byurakan Astrophysical Observatory}{}{Byurakan}{}{}{Armenia}
\paperauthor{M.~Turatto}{}{}{Osservatorio Astronomico di Padova}{}{Padova}{}{}{Italy}

\begin{abstract}
  We present the summary of our last results on the spatial distribution and relative frequencies of Supernovae (SNe)
  in a large number of host galaxies from the Sloan Digital Sky Survey (SDSS).
  We use the locations of SNe in order to study the relations between radial/azimuthal distributions of SNe
  and properties of their hosts and environments.
  On the other hand, the vertical distribution of SNe allows to study the progenitors association to
  the thin or thick discs, and to the stellar halo.
  We also propose the underlying mechanisms shaping the number ratios of SNe types.
  It is important to note that there were no extended studies of the 3D distribution of SNe and
  structural parameters of hosts. Our study is intended to fill this gap and
  better constrain the nature of SN progenitors.
\end{abstract}

\section{Introduction}

A crucial aspect of many recent studies of extragalactic SNe
is to establish the links between the nature of SN progenitors and
stellar populations of their host galaxies.
The most direct method for this is through their identification on pre-SN images.
However, the number of such SNe is small and is limited to the nearby core-collapse (CC)
events \citep[e.g.][]{2009ARA&A..47...63S}.\footnote{{\footnotesize SNe are
generally divided into two categories according to their progenitors:
CC and Type Ia SNe. CC SNe result from massive young stars that undergo CC
\citep[e.g.][]{2003LNP...598...21T,2012MNRAS.424.1372A},
while Type Ia SNe are the end point in the evolution of binary stars when
an older white dwarf (WD) accretes material from its companion,
causing the WD mass to exceed the Chandrasekhar limit,
or a double WD system loses angular momentum due to gravitational wave
emission, leading to coalescence and explosion
\citep[e.g.][]{2014ARA&A..52..107M}.}}
These limitations force us to study the
properties of SN progenitors through indirect methods.

The properties of SN host galaxies, such as the morphology, color, nuclear activity,
star formation rate (SFR), metallicity, stellar population, age, etc.
provide strong clues to the understanding of the progenitors
\citep[e.g.][]{2005A&A...433..807M,2005AJ....129.1369P,2008A&A...488..523H,2009A&A...503..137B,
2010ApJ...721..777A,2012ApJ...759..107K,2012MNRAS.424.1372A,2015MNRAS.448..732A,2016MNRAS.459.3130A}.
SN hosts in multiple systems of galaxies can also provide important constraints
relating the star formation distribution/rate in strongly interacting systems
with the different SN progenitors \citep[e.g.][]{2011MNRAS.416..567A,2012A&A...540L...5H}.
In particular, valuable information of the nature of progenitors can be obtained through
the study of the radial \citep[e.g.][]{2008Ap.....51...69H,2008MNRAS.388L..74F,
2009A&A...508.1259H,2013MNRAS.436.3464K},
azimuthal \citep[e.g.][]{1995A&A...297...49P,2001MNRAS.328.1181N},
and vertical \citep[e.g.][]{2016AstL...42..495P} distributions of SNe and their environments
\citep[e.g.][]{2011A&A...530A..95L,2016A&A...591A..48G}.

By this contribution, we present a brief summary of our results recently obtained in this field.

\section{The database}

\citet[][]{2012A&A...544A..81H} provides a large and well-defined database
that combines extensive new measurements and a literature search of about four thousand SNe
and their host galaxies located in the sky area covered by the SDSS
Data Release 8. This database is much larger than previous ones, and
provides a homogeneous set of global parameters of SN hosts,
including measurements of apparent magnitudes, diameters, axial ratios, position angles,
morphological classifications, and activity classes of nuclei.
Special attention was paid to collect accurate data on the spectroscopic classes,
coordinates, offsets of SNe, and heliocentric redshifts of the host galaxies.
In \citet[][]{2014MNRAS.444.2428H}, we also classified the morphological disturbances
of nearby host galaxies from the visible signs of galaxy--galaxy interactions
in the SDSS.\footnote{{\footnotesize The levels of disturbance are arranged
in an approximate chronological order according to the different stages of interaction.}}

All the presented results in the next section are mostly based on this database,
which is publicly available.

\section{Summary of our results}

In this section, we list the results accompanied with short discussions.
For more details, the reader is referred to
\citet[][]{2012A&A...544A..81H,2014MNRAS.444.2428H,2016MNRAS.456.2848H,2013Ap&SS.347..365N}.

\subsection*{The number ratio of SN types}

\begin{itemize}
\item The mean morphological type of spiral galaxies hosting Type Ia SNe is significantly earlier
      than the mean host type for all other types of CC SNe,
      which are consistent with one another.\footnote{{\footnotesize CC SNe
      are observationally classified in three major classes, according to
      the strength of lines in optical spectra \citep[e.g.][]{1997ARA&A..35..309F}:
      Type II SNe show hydrogen lines in
      their spectra, including the IIn (dominated by emission lines with narrow components)
      and IIb (transitional objects with observed properties closer to SNe II at early times,
      then metamorphosing to SNe Ib) subclasses;
      Type Ib SNe show helium but not hydrogen, while Type Ic SNe show neither hydrogen nor helium.
      All these SNe types arise from young massive progenitors with possible differences in
      their masses, metallicities, ages, and fractions of binary stellar systems
      \citep[e.g.][]{2011MNRAS.412.1522S}.}}
\item We find a strong trend in the behaviour of $N_{\rm Ia}/N_{\rm CC}$ depending on
      host-galaxy morphological type, such that early-type (high-mass or high-luminosity)
      spirals include proportionally more Type Ia SNe.\footnote{{\footnotesize The
      $N_{\rm Ia}/N_{\rm CC}$ ratio is a good tracer of specific SFR
      [sSFR] \citep[e.g.][]{2009A&A...503..137B}.}}
      The behaviour of $N_{\rm Ia}/N_{\rm CC}$ versus morphology is a simple
      reflection of the behaviour of 1/sSFR versus morphological types of galaxies
      \citep[e.g.][]{2009A&A...503..137B}.
\item The $N_{\rm Ia}/N_{\rm CC}$ ratio is nearly constant when changing
      from normal, perturbed to interacting galaxies, then declines in merging galaxies, whereas it
      jumps to the highest value in post-merging/remnant host galaxies.
      During the relatively short time-scale of the merging stage \citep[e.g.][]{2008MNRAS.391.1137L},
      the spiral, gas-rich galaxies do not have enough time to produce many Type Ia SNe,
      but can intensively produce CC SNe, assuming short lifetimes for the CC SNe progenitors.
      In the post-merging/remnant galaxies with longer time-scale,
      the SFRs and morphologies of host galaxies are strongly affected,
      significantly increasing the $N_{\rm Ia}/N_{\rm CC}$ ratio.
\item The $N_{\rm Ibc}/N_{\rm II}$ ratio is nearly constant when changing
      from normal, perturbed to interacting galaxies,
      then jumps to the highest value in merging galaxies
      and slightly declines in post-merging/remnant subsample.\footnote{{\footnotesize By SN~Ibc,
      we denote stripped-envelope SNe of Types Ib and Ic, as well as mixed Ib/c
      whose specific subclassification is uncertain.}}
      In our merging hosts, the positions of CC SNe, particularly SNe of Ibc type, mostly coincide
      with the circumnuclear regions and only in few cases with bright H~{\footnotesize II} regions,
      which is in agreement with the previously found central excess of CC SNe in
      extremely disturbed or merging galaxies \citep[e.g.][]{2012MNRAS.424.2841H,2012A&A...540L...5H}.
\item Type Ibc SNe are located in pairs of galaxies with significantly smaller difference of
      radial velocities between components than pairs containing Types Ia and II SNe.
      We consider this as a result of higher SFR of these closer systems of galaxies
      \citep[e.g.][]{2011MNRAS.412..591P}.
      SN types are not correlated with the luminosity ratio of
      host and neighbor galaxies in pairs.
\item The $N_{\rm Ia}/N_{\rm CC}$ ($N_{\rm Ibc}/N_{\rm II}$) ratio increases (decreases)
      when moving from SF, C, to Sy$+$LINER activity classes for the host galaxies. In the invoked scenario,
      the interaction is responsible for morphological disturbances and for partially sending gas inward,
      which first triggers star formation \citep[e.g.][]{2013MNRAS.430..638S}
      and increases sSFR \citep[e.g.][]{2008MNRAS.391.1137L}.
      Therefore, in the SF stage, we observe a lower value of
      the $N_{\rm Ia}/N_{\rm CC}$ ratio and a somewhat higher value of
      the $N_{\rm Ibc}/N_{\rm II}$ ratio as in morphologically disturbed (interacting or merging)
      late-type galaxies. The starburst then fades with time and the C (composite of SF and AGN) class evolves to
      the AGN [Sy$+$LINER] \citep[e.g.][]{2010MNRAS.405..933W} with a comparatively relaxed disturbance,
      early-type morphology, poor gas fraction, and old stellar population. Therefore,
      in the AGN stage, we observe inverse values of the ratios as in morphologically
      less disturbed (relaxed) early-type galaxies.
\end{itemize}

\subsection*{The radial distribution of SNe}

\begin{itemize}
\item In Sa--Sm galaxies, all CC and the vast majority of Type Ia SNe belong to the disc,
      rather than the bulge component.
      This result suggests that the rate of SNe Ia in spiral galaxies is dominated by
      a relatively young/intermediate progenitor population
      \citep[e.g.][]{2005A&A...433..807M,2011Ap.....54..301H,2011MNRAS.412.1473L}.
\item The radial distribution of Type Ia SNe in S0--S0/a galaxies is inconsistent
      with that in Sa--Sm hosts.
      This inconsistency is mostly attributed to the contribution
      by SNe Ia in the outer bulges of S0--S0/a galaxies.
      In these hosts, the relative fraction of bulge to disc SNe Ia is probably changed
      in comparison with that in Sa--Sm hosts, because the progenitor population from
      the discs of S0--S0/a galaxies should be much lower due to the lower number of
      young/intermediate stellar populations.
\item The radial distribution of CC SNe in barred Sa--Sbc galaxies is not consistent
      with that of unbarred Sa--Sbc hosts,
      while for Type Ia SNe the distributions are
      not significantly different.
      At the same time, the radial distributions of both
      Type Ia and CC SNe in Sc--Sm galaxies are not affected by bars.
      These results are explained by a substantial suppression of
      massive star formation in the radial range swept by strong bars of
      early-type barred galaxies \citep[e.g.][]{2009A&A...501..207J,2015MNRAS.450.3503J}.
\item The radial distribution of CC SNe, in contrast to Type Ia SNe, is inconsistent with
      the exponential surface density profile, because of the central ($R_{\rm SN}/R_{25} \lesssim 0.2$)
      deficit of SNe.\footnote{{\footnotesize The $R_{25}$ is the SDSS $g$-band
      $25^{\rm th}$ magnitude isophotal semimajor axis of SN host galaxy.}}
      However, in the $R_{\rm SN}/R_{25}\in[0.2; \infty)$ range, the inconsistency
      vanishes for CC SNe in most of the subsamples of spiral galaxies.
      In the inner-truncated disc, only the radial distribution of CC SNe in barred early-type spirals
      is inconsistent with an exponential surface density profile, which
      appears to be caused by the impact of bars on the radial distribution of CC SNe.
\item In the inner regions of non-disturbed spiral hosts, we do not detect
      a relative deficiency of Type Ia SNe
      with respect to CC, contrary to what was found by other authors
      \citep[e.g.][]{1997ApJ...483L..29W,2015MNRAS.448..732A}, who
      had explained this by possibly stronger dust extinction for Type Ia than for CC SNe.
      Instead, the radial distributions of both types of SNe are similar
      in all the subsamples of Sa--Sbc and Sc--Sm galaxies, which supports
      the idea that the relative increase of CC SNe in the inner regions of spirals found by the other
      authors is most probably due to the central excess of CC SNe in disturbed galaxies
      \citep[e.g.][]{2012MNRAS.424.2841H,2012A&A...540L...5H}.
\end{itemize}

\subsection*{The azimuthal and vertical distributions of SNe}

\begin{itemize}
\item The orientation of SNe in host with respect to the preferred direction
      toward neighbor galaxy is found to be isotropic and independent of kinematical
      properties of the galaxy pair.
\item The vertical distribution of CC SNe is significantly different from that of Type Ia SNe,
      being about twice more concentrated to the plane of the disc in edge-on host galaxies.
      The vertical distribution of CC SNe can be assigned to the younger thin disc population,
      while the distribution of Type Ia SNe is consistent with the older thick disc of edge-on galaxies.
\end{itemize}

The obtained results show that the spatial distributions and the number ratios of different SNe
are powerful tools to constrain the natures of their progenitors and
to better understand the star formation processes in various types of galaxies.

{\footnotesize \acknowledgements This work was supported by the RA MES State Committee of Science,
in the frames of the research project number 15T--1C129.
This work was made possible in part by a research grant from the
Armenian National Science and Education Fund (ANSEF) based in New York, USA.
V.A. acknowledges the support from FCT through Investigador FCT contracts of reference IF/00650/2015.
V.A. also acknowledges the support from Funda\c{c}\~ao para a Ci\^encia e Tecnologia (FCT) through national funds and from FEDER through COMPETE2020 by the following grants UID/FIS/04434/2013 \& POCI-01-0145-FEDER-007672, PTDC/FIS-AST/7073/2014 \& POCI-01-0145-FEDER-016880 and PTDC/FIS-AST/1526/2014 \& POCI-01-0145-FEDER-016886.}



\end{document}